\documentclass[useAMS,usenatbib]{mn2e}
\usepackage{graphicx}
\usepackage{rotating} 
\usepackage{longtable}
\usepackage{lscape}
\usepackage[usenames]{color}

\def\lsim{\mathrel{\rlap{\lower 3pt \hbox{$\sim$}} \raise 2.0pt \hbox{$<$}}}
\def\gsim{\mathrel{\rlap{\lower 3pt \hbox{$\sim$}} \raise 2.0pt \hbox{$>$}}}
\def\nat{Nat}
\def\apj{ApJ}
\def\apjl{ApJL}

\def\mnras{MNRAS}

\def\aj{AJ}

\def\pasp{PASP}

\def\msun{\rm M_{\odot}}

\title[Observational signatures of sub-pc binary black
holes II]{Search of sub-parsec massive binary black holes through
line diagnosis II}

\author[Montuori et
al.]{C. Montuori$^{1}$\thanks{carmen.montuori@uninsubria.it},
  M. Dotti$^{2}$, F. Haardt$^{1,3}$, M. Colpi$^{2}$, R. Decarli$^{4}$ \\
             $^1$ Universit\`a degli Studi dell'Insubria, Via Valleggio 11,
             22100 Como, Italy\\
             $^2$ Universit\`a degli Studi di Milano-Bicocca, Piazza della
             Scienza 3, 20126 Milano, Italy\\
              $^3$ INFN, Sezione di Milano-Bicocca, Piazza della
              Scienza 3, 20126 Milano, Italy\\
              $^4$ Max Planck Institut f\"{u}r Astronomie, K\"{o}nigstuhl 17,
             69117 Heidelberg, Germany}
            
\pagerange{\pageref{firstpage}--\pageref{page}} \pubyear{201?}

\begin{document}        
\maketitle
\label{firstpage}

\begin{abstract}
  Massive black hole binaries at sub-parsec separations may display in
  their spectra anomalously small flux ratios between the MgII and CIV
  broad emission lines, i.e. $F_{\rm{MgII}}/F_{\rm{CIV}}\lsim 0.1$,
  due to the erosion of the broad line region around the active,
  secondary black hole, by the tidal field of the primary.  In Paper I
  by Montuori et al. (2011), we focussed on broad lines emitted by gas
  bound to the lighter accreting member of a binary when the binary is
  at the center of a hollow density region (the gap) inside a
  circum-binary disc. The main aim of this new study is at exploring
  the potential contribution to the broad line emission by the
  circum-binary disc and by gaseous streams flowing toward the black
  hole through the gap. We carry out a post-process analysis of data
  extracted from a SPH simulation of a circum-binary disc around a
  black hole binary. Our main result is that the MgII to CIV flux
  ratio can be reduced to $\sim 0.1$ within an interval of sub-pc
  binary separations of the order of $a \sim (0.01-0.2)
  (f_{\rm{Edd}}/0.1)^{1/2}$ pc corresponding to orbital periods of
  $\sim (20-200) (f_{\rm{Edd}}/0.1)^{3/4} $ years for a secondary BH mass in the range $M_2 \sim
  10^7-10^9 \, \rm{M_{\odot}}$ and a binary mass ratio of 0.3. At even
  closer separations this ratio returns to increase to values that are
  indistinguishable from the case of a single AGN (typically $F_{\rm{MgII}}/F_{\rm{CIV}}\sim 0.3-0.4$) because of the
  contribution to the MgII line from gas in the circum-binary disc.
  
\end{abstract}
\begin{keywords}
black hole physics -- galaxies: kinematics and dynamics -- galaxies: nuclei --
quasars 
\end{keywords}
\section{Introduction}\label{sec:intro}
Recent progress on the observational search of Active Galactic Nuclei
(AGN) with peculiar features led to the discovery of a sizable number
of active black hole (BH) pairs in galaxies undergoing a merger,
dubbed dual AGN. These pairs are observed when the two galaxies are
in the early stage of a merger, at separations in excess of several
kpc (e.g. Comerford et al. 2009, Green et al. 2010, Liu et al. 2010,
Fu et al. 2011, Shields et al. 2012). The dual BHs are expected to evolve into a
close pair, as they sink in the gravitational potential of the new
galaxy.  At about a few parsec the BHs form a Keplerian binary, and
this occurs when their mass exceeds the mass of stars and gas enclosed
in their orbit (Colpi \& Dotti 2011 for a review).  This phase and the
phase of subsequent hardening due to the interaction of the
stellar/gaseous background is still difficult to observe. At such
close separations the binary (if both BHs are active) would not be
identified as a dual source due to insufficient spatial resolution.

The discovery of bona fide BH binaries (BHBs hereon) is important as
it represents a fundamental test for current models of galaxy
formation: BHBs are signposts of galactic mergers and thus of the
cosmic assembly of substructures.  Nonetheless, convincing
observational evidence of close BHBs in galactic nuclei is still
missing with few exceptions.  Two compact double radio nuclei are
observed in the core of the galaxy 0402+379 on scales of $\sim 7$ pc
(Rodriguez et al. 2006), and signatures of a Keplerian motion of a
sub-pc BH binary have been claimed in OJ287 (Valtonen et al. 2008 and references therein).  

Spectroscopic search of BHBs is a promising technique for the search
of such elusive sources. The BHs in a close binary revolve around the
common center of mass with a velocity that can highly exceed the
velocity of gas and stars far out in the gravitational potential of
the background galaxy, and this may generate a velocity off-set among
atomic emission lines. Taking advantage of the wealth of
spectroscopic data made publicly available thanks to the Sloan Sky
Digital Survey (SDSS, York et al. 2000), different groups developed systematic strategies
focussed on the selection of AGN spectra displaying multiple sets of
narrow and broad emission lines (BELs) with relative Doppler-shifts
$\gsim 1000 \,\rm{km \, s^{-1}}$ (e.g. Tsalmantza et al. 2011, 
Eracleous et al. 2011, Tsalmantza \& Hogg 2012). Substantial velocity off-sets
between the two set of emission lines, respectively associated with
the host galaxy reference frame and the gas bound to the single BH,
potentially mark the orbital motion of the active member of the binary
system in the galactic nucleus.  Follow-up campaigns revealing
periodic changes in the observed Doppler-shifted lines are necessary
in order to firmly favor the binary hypothesis against other possible
physical scenarios such as: a recoiling BH resulting from a BHB
coalescence, a double-peaked emitter or a chance superposition along
the line of sight. Considering that the orbital period of a sub-pc BHB
can be as long as a few $100$ yr, the spectral monitoring of these
sources is rather long. A further disadvantage of this direct
Doppler-shift technique is that the spectroscopic search is limited to
a redshift of $z \lsim 0.8$ as at higher redshifts all the most
important narrow emission lines fall out of the optical range covered
by the survey.

In Montuori et al. (2011; hereafter Paper I), we devised a tool for exploring
potential spectral signatures of sub-pc BHBs at redshift $z\sim 2$, where BHBs
should be frequent according to the present hierarchical cosmological
model. The explored scenario was that of a binary, surrounded by a circum-binary
disc, that after clearing a gap, i.e. a low density hollow region in the midst
of the disc, has confined its activity around the secondary BH fed by material
inflowing from the inner edge of the disc (e.g. Artymovicz \& Lubow 1994,
Ivanov et al. 1999, Hayasaki et al. 2007, 2008, Cuadra et al. 2009). We
assumed that a small-scale accretion disc is maintained around the secondary
BH throughout the binary orbital decay induced by the external torques from
the circum-binary disc or by the emission of gravitational waves
(e.g. Haiman, Kocsis \& Menou 2009). In Paper I we noticed that under these conditions broad
emission line clouds orbiting around the active BH suffer erosion inside the
gap due to the tidal truncation at the Roche Lobe surface, resulting in a
sizable reduction of the flux ratio between lines of low and high ionization
potentials relative to the typical values observed for isolated AGN. In
particular it was found that sub-parsec BHBs could be potentially identified
in the archives of current automatic surveys selecting AGN spectra
characterized by a flux ratio between the MgII and the CIV lines
($F_{\rm{MgII}}/F_{\rm{CIV}}$) below $\lsim 0.1$.

In Paper I, the contribution to the different BELs from illuminated
gas not bound to the secondary, active BH was neglected.  In this
paper, we improve upon our earlier investigation by studying the
contribution to the BELs resulting from reprocessing of BH radiation
by the gas of the circum-binary disc and by gas streams in the gap
region flowing toward the BHB.  In particular we explore when and how
the ratio $F_{\rm{MgII}}/F_{\rm{CIV}}$ varies in response to changes
in the spatial distribution of the broad line emitting gas as the binary hardens.

Section \ref{sec:qualitative} and
\ref{sec:quantitative} describe how we extend our previous result, 
guided by a qualitative analysis, and later by a more
quantitative study.  In Section \ref{sec:results} we present
the results of this analysis, carried out with data taken from SPH
numerical simulations. Discussion and conclusions are reported in
Section \ref{sec:discussion}.

\section{Key spatial scales along binary evolution}\label{sec:qualitative}

\begin{figure}
\centering
\includegraphics [scale=0.37] {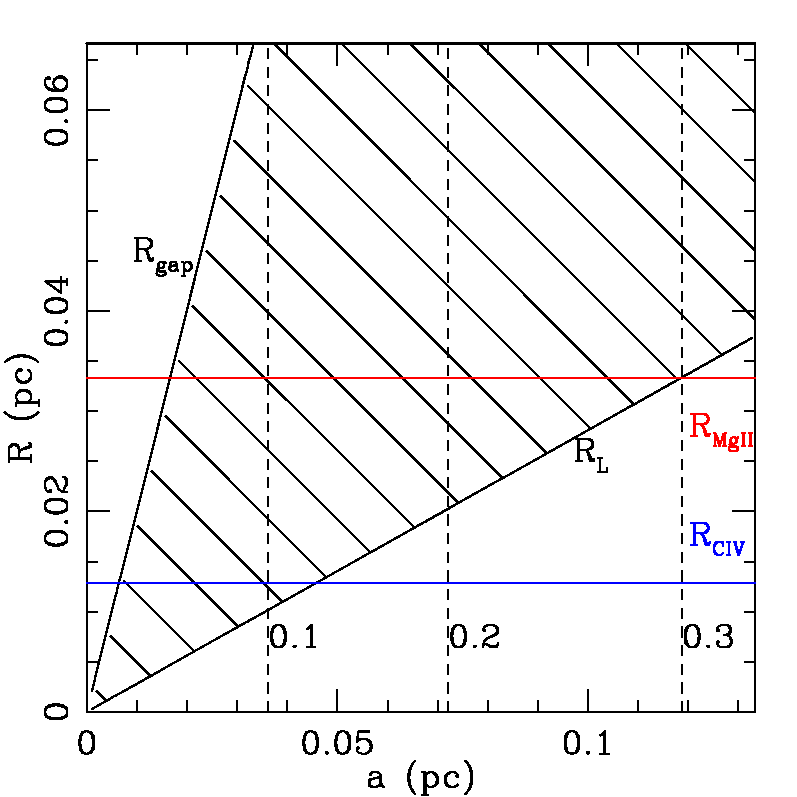}
\caption{\small{Schematic view of the most relevant
    spatial scales in our BHB model as a function of $a$. The 
    binary is interacting with a circum-binary disc,
    and the secondary BH is the only active member of the system. The
    binary mass ratio is $q=0.3$, the secondary BH mass is 
    $M_2=10^8 \,\msun$. The two solid black lines represent the radii of the
    Roche lobe ($R_{\rm L}$) around the secondary BH, and the radius of the inner edge of the
    circum-binary disc ($R_{\rm gap}$), respectively.  The shaded area in between these two
    lines corresponds to the lower density region in the central part
    of the circum-binary disc, i.e. the gap. The red and blue
    horizontal lines correspond to the values of $R_{\rm BLR}$
    computed from the radius-luminosity relationship for the MgII and the CIV
    emission lines, respectively. The secondary BH is assumed to be
    active as an AGN with an Eddington ratio of 0.1. The vertical
    dashed lines mark the orbital separations corresponding to
    different values of  $F_{\rm{MgII}}/F_{\rm{CIV}}$ 
    computed for the material bound to the active BH, orbiting inside
    $R_{\rm{L}}$ (see the text for more details).} }
\label{fig:radii}
\end{figure}

In order to overcome some of the simplifying assumptions made in our
exploratory Paper I, we consider here the potentially observable
contribution from gas in the circum-binary disc surrounding the BHB.
We start with a schematic view of the spatial scales relevant to our
binary evolutionary model.  Two characteristic radii are present: the
Roche Lobe Radius $R_{\rm L}$ around the secondary, only active BH and
the inner edge of the circum-binary disc $R_{\rm gap}$ . The Roche
Radius $R_{\rm L}\sim 0.49 \, a \, q^{2/3} /[0.6q^{2/3} +\ln(1 + q^{1/3})]$
(Eggleton 1983)
and $R_{\rm gap}=2a$ (e.g. Artymovicz \& Lubow 1994) are plotted in Figure \ref{fig:radii} as function
of the binary orbital separation $a$, for a circular BHB with mass
ratio $q=0.3$ and mass of the secondary $M_2=10^8\,\msun$.  The shaded
area corresponds to the gap region between $R_{\rm L}$ and $R_{\rm gap}$,
while the horizontal lines correspond to the values of $R_{\rm{BLR}}$
calculated from the following radius-luminosity relations for the CIV
($R_{\rm CIV}$, in blue, Kaspi et al. 2007) and the MgII ($R_{\rm
  MgII}$, in red, McLure \& Jarvis 2002): 
\begin{equation}
R_{\rm{CIV}} = 0.014 \times \left ( \lambda L_{1350}/10^{43}
    \rm{erg\,s^{-1}} \right )^{0.55} \, \rm{pc}
\end{equation}
\begin{equation}
R_{\rm{MgII}} = 0.022 \times \left ( \lambda
    L_{3000}/10^{44}\rm{erg\,s^{-1}}\right )^{0.47} \, \rm{pc},
\end{equation}
where $\lambda L_{1350}$ and $\lambda L_{3000}$ are the continuum luminosities
at 1350$\AA$ and 3000$\AA$. We consider the case of a secondary BH shining
with a constant Eddington ratio set to $f_{\rm Edd}=0.1$. The vertical dotted
lines mark the orbital separations corresponding to different values of
$F_{\rm MgII}/F_{\rm CIV}$ (0.1, 0.2 and 0.3 as indicated in the figure) for
the lines emitted by the gas of the BLR located inside $R_{\rm L}$. These
orbital separations are computed as described in Paper I. In particular, the
emission from the BLR around the secondary BH is modelled according to the LOC
model (e.g. Baldwin et al. 1995), adopting a power-law with index set to -1
for the clouds radial distance from the active source, and a uniform
distribution for the clouds density at each radius. The hydrogen number
density is set in the range $10^9 {\rm{cm^{-3}}}\le n_{\rm H} \le 6 \times 10^{12} \, \rm
cm^{-3}$ so that at $a_0\simeq 0.12$ pc, our starting point where $R_{\rm L}$
coincides with $R_{\rm{MgII}}$,  $F_{\rm MgII}/F_{\rm CIV}$ is consistent with the typical
values observed for standard AGN, i.e. $\sim 0.3 - 0.4$ (see figure 4 of Paper
I). Moving toward smaller BHB separations, we highlight the occurrence of
intervals possibly characterized by different observational signatures. As the
binary separation shrinks, $R_{\rm{L}}$ becomes smaller
than $R_{\rm MgII}$. According to our previous results, this implies a
reduction in the ratio $F_{\rm MgII}/F_{\rm CIV}$ for the BELs associated with
the BLR of the secondary BH. We consider the results of Paper I to be valid
until the flux ratio drops to a value $F_{\rm MgII}/F_{\rm CIV}\sim 0.1$ at an
orbital separation of $a \simeq 0.04$ pc, corresponding to a period of $P \simeq
40$ yr, for the selected binary parameters.  The contribution coming from the
material not bound to the secondary active BH can be still neglected at these
separations considering that: (i) the gas density in the gap region is
expected to be much lower than the density relative to standard BLRs, and the
BELs are efficiently reprocessed only at high densities; and that (ii) the
inner edge of the circum-binary disc $R_{\rm gap}$ is located at a distance $>
2 R_{\rm MgII}$ so that the ionizing flux reaching the higher density material
associated with the outer circum-binary disc would be more than an order of
magnitude lower relative to the flux intercepted by the gas bound to the
active source.

As the semimajor axis shortens further on, this spectroscopic signature is not
expected to be a good tracer of the presence of a sub-pc BHB, 
since the additional contribution of the circum-binary disc can not be any longer neglected,
yielding an increase in $F_{\rm MgII}/F_{\rm CIV}$. At 
intermediate separations of the order of $ a \simeq 0.02$ pc, where $R_{\rm gap}
\approx R_{\rm MgII}$, $F_{\rm MgII}/F_{\rm CIV}$ could be even higher than what
typically observed for isolated AGN since the circum-binary disc could
start to contribute significantly to the MgII line while the CIV
emission flux could be still reduced relative to the case of a single
AGN since $R_{\rm L} < R_{\rm CIV} < R_{\rm gap}$. At the smallest orbital
separations, of the order of $a < 0.01$ pc, where
$R_{\rm gap} \lsim R_{\rm CIV}$, the higher density gas of the circum-binary
disc could efficiently reprocess both the CIV and the MgII emission
lines bringing $F_{\rm MgII}/F_{\rm CIV}$ back to the values observed for standard
BLRs.

\section{Reprocessing of radiation in simulated circum-binary discs}\label{sec:quantitative}

In Section \ref{sec:qualitative}
we qualitatively showed that  the flux ratio of the MgII and CVI emission lines varies with 
BHB separation, and that this change may not be monotonic with
decreasing $a,$ due to the contribution to the line flux from dense gas 
present in the circum-binary disc, depending on the
relative position of the two critical radii $R_{\rm CVI}$ and $R_{\rm MgII}$
with respect to $R_{\rm gap}$ and $R_{\rm L}$.

To proceed more quantitatively in our analysis, we model the spectroscopic signatures of a close BHB
applying here a simple line radiative transfer algorithm
on data extracted from a simulated model
of a circum-binary gas disc. The model self-consistently accounts for
the presence of a low (but finite) density region, i.e. the gap, and
of gas streams that leaking out through the inner edge of
the disc are accreted by the binary. 
 
In particular we consider the numerical results presented in Sesana et
al. (2012, S12 hereafter) who used a modified version of the SPH code
Gadget-2 (Springel 2005, Cuadra et al. 2009) to model the complex
feeding process of a BHB embedded in a coplanar, co-rotating
circum-binary accretion disc.  The orbital separation in the
simulation, $a\simeq 0.01$ pc, is in the interval appropriate to test
our working hypothesis, namely that the line fluxes from MgII and CVI
can have a contribution from dense gas in the circum-binary disc.  At
the considered orbital separation, the simulated binary system is
still dynamically coupled with the circum-binary disc, while its
orbital decay is dominated by the emission of gravitational waves (see
S12 and references therein).

 We map the gas density field on a cubic, three
dimensional coarse grid\footnote {The results of our analysis are
  not significantly affected by the choice of the grid geometry and
  resolution.}, and in the following we will refer to the cubic grid
  elements as pixels. Pixels are then exposed to the ionizing
radiation originating from gas accreting in the putative small-scale
disc around the secondary BH.
The small-scale accretion disc is assumed to be coplanar with the binary, and the
ionizing emission is modeled adopting a cosine-like dilution along the
disc axis to mimic thermal emission from an optically thick (geometrically thin)  accretion
flow (e.g. Ivanov et al 1999, Shakura \& Sunyaev 1973). In the SPH simulation at hand the gas accretion rate onto the BH
is given by summing over all particles crossing the BH sink radius in
a time-step. In our analysis, we select simulation snapshots
corresponding to the secondary BH radiating with a luminosity of 0.1
the Eddington luminosity (assuming 10\% radiative efficiency).
Then, for all simulation pixels, 
the gas ionization state is computed using the code CLOUDY (version 08.00; Ferland et al. 1998).  
We adopt the AGN SED template stored in the code, scaled to the appropriate level given by the 
coordinates of the considered pixel. 
The ionization state of the gas is determined moving from 
the innermost pixels to the outer region along straight ray paths. 
To trace the cumulative effect of radiation absorption along a ray path, 
we make a net distinction between pixels which are optically thick and optically
thin to the HI ionizing continuum. We assume that a pixel is not reached by ionizing radiation 
if at least one optically thick pixel is met along a ray. Our treatment is clearly a simplified
description of the real radiative transfer, as we assume complete (zero) absorption for $\tau \ge 1$
($\tau < 1$). However this procedure greatly
facilitates the analysis process, and at the same time captures the basic feature of sharp ionization fronts 
expected in the circum-binary gas.

The next step is to model the BELs. A self-consistent model for the
BLR is still lacking as it is poorly constrained by the
observations. Despite these limitations, the origin of the BELs is
generally attributed to compact, dense condensations rapidly moving in
a hotter intercloud medium (e.g. Osterbrock \& Ferland 2006).  Though the density of such unresolved
clouds (estimated to be $\gsim 10^{10} \, \rm{cm^{-3}}$ ) is well
above the maximum density given by our SPH simulation, we can use the
gas distribution determined in the SPH simulation as a physically
motivated "background" density field to model the spatial distribution
of the BEL clouds. To this aim, we select the densest ($n_{\rm H} >
10^8 \, \rm{cm^{-3}}$) ionized regions of the circum-binary disc and
of the streams flowing from its inner-edge, and there we superimpose
a distribution of gas overdensities mimicking the BEL clouds.
Operatively, we add a gaseous clump in all illuminated pixels whose
hydrogen number density is above a given threshold. Such limiting
value is set so that the total number of pixels embedding BEL clumps
do not exceed the $10 \%$ of the total number of ionized pixels. The
density of each BEL clump is set randomly from a uniform distribution
in the range $10^8 \, \rm{cm^{-3}}\le n_{\rm H }\le 10^{12} \,
  \rm{cm^{-3}}$.  We further assume that the clumps filling factor is $\ll
1$, so that the cumulative shielding effect along ray paths can be
neglected. As discussed in the following section, we manage to check
the consistency of our expectations discussed in Section
\ref{sec:qualitative} without any other constraint on the
distribution of such BEL clumps.

Once BEL clumps are added, we use again CLOUDY to compute
their line emission properties. As a final
remark, we caution that we do not model the emission coming
from the gas orbiting inside the Roche lobe of the secondary. 
The numerical resolution is not high enough to describe the properties
of the material located at radii $\ll 0.5 \, a$. We note that
the main interest of our work here is to compute the broad line
emission from the material not bound to the active BH but rather bound to the binary, in order to
test our qualitative expectations.

\section{Results}\label{sec:results}

\begin{figure*}
\centering
\includegraphics [scale=0.3] {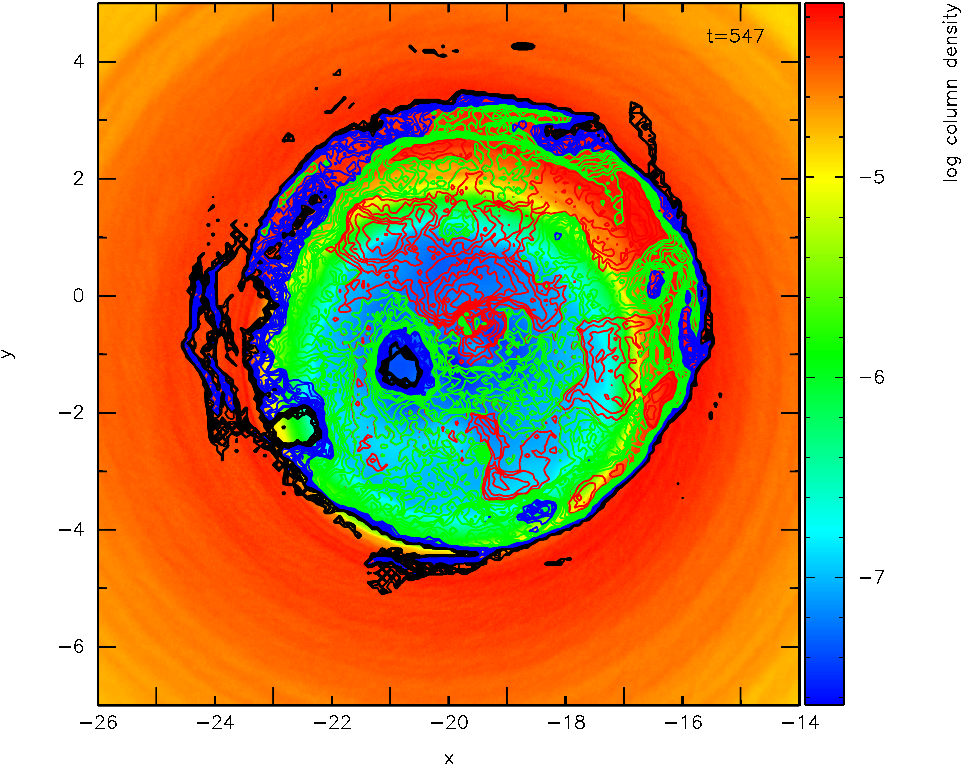}
\includegraphics [scale=0.3] {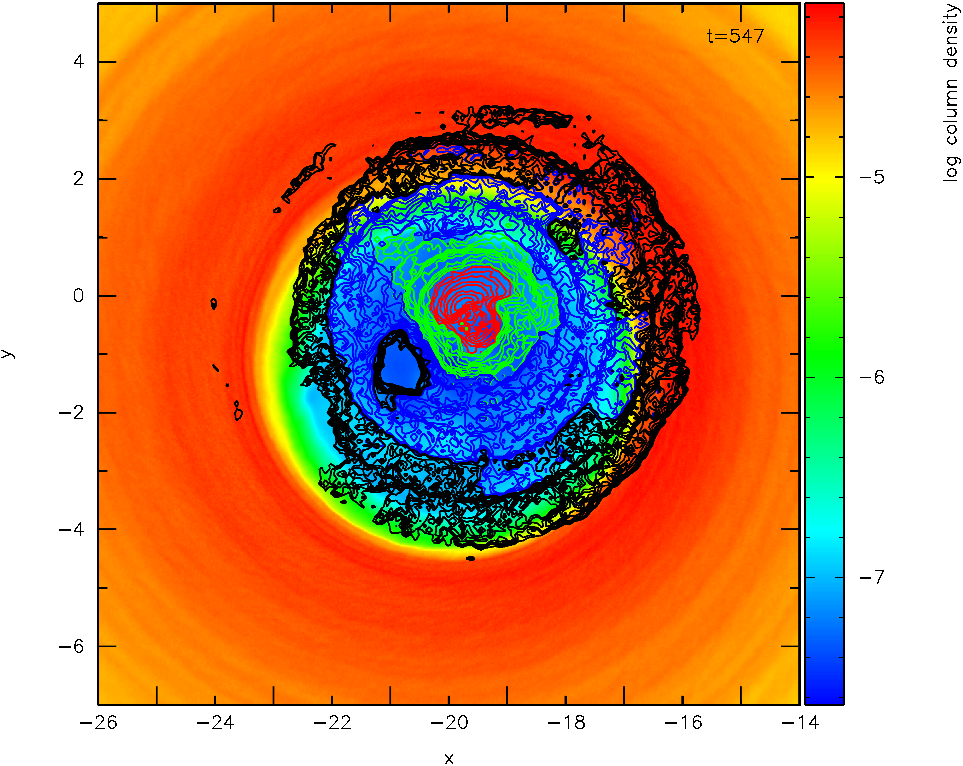}
\includegraphics [scale=0.34] {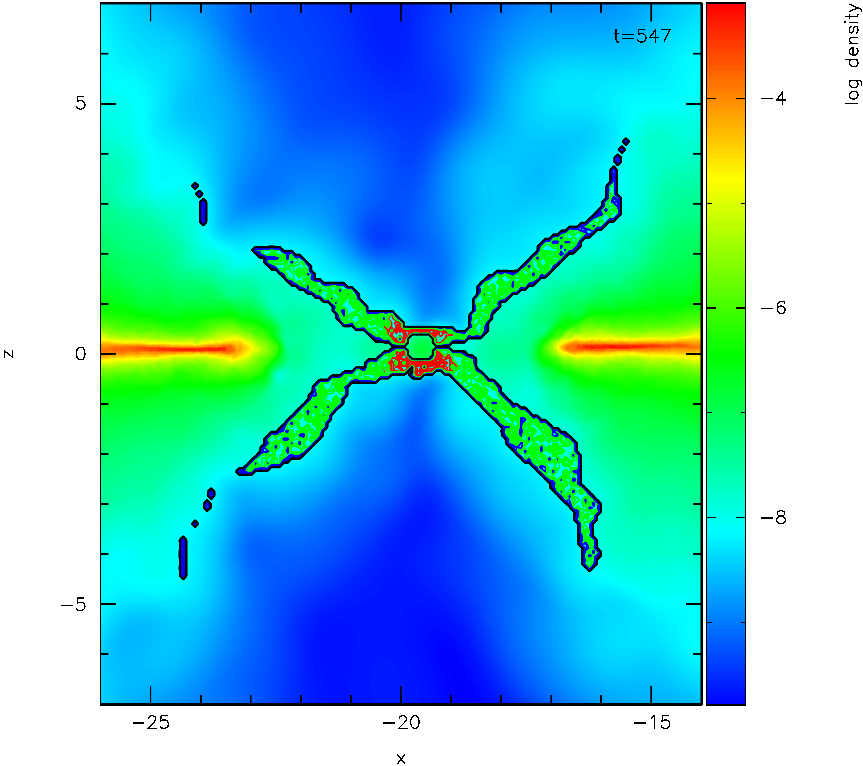}
\includegraphics [scale=0.34] {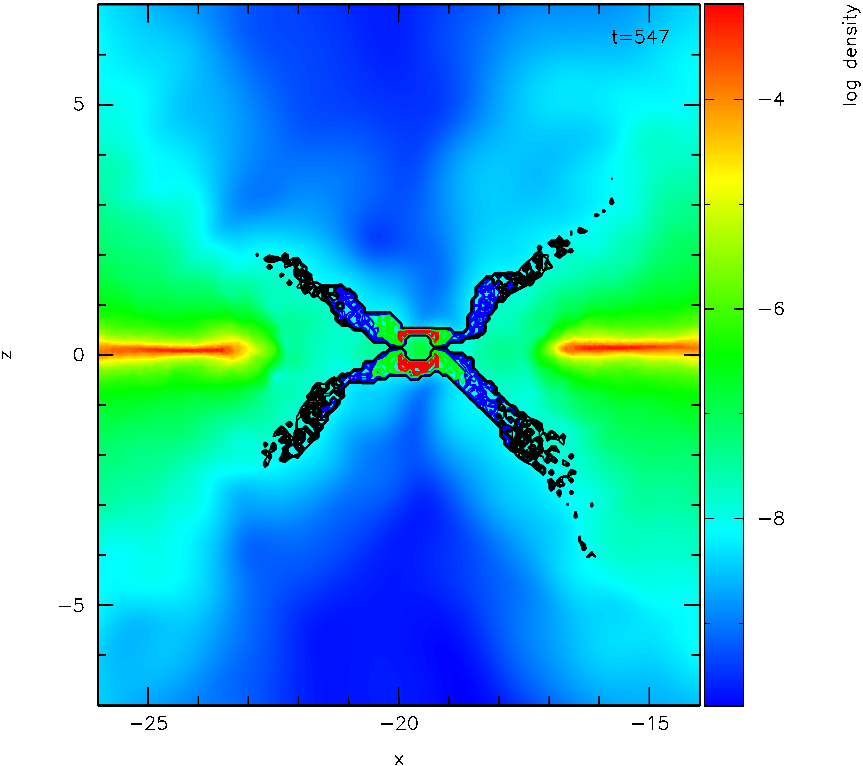}
\caption{\small { Contour plots of the MgII (left panels) and CIV
    (right panels) emission line intensity superimposed on the density
    map of a snapshot taken from a SPH numerical simulation of a
    massive BHB embedded in a coplanar, co-rotating circum-binary
    disc. The upper panels show the projected intensity map for a face
    on view relative to binary orbital plane superimposed on the
    surface density map.  The lower panels correspond to an edge-on
    view with a slice of the intensity map overlaid on a slice of the
    density map, both taken in the $z-x$ plane at the $y$-coordinate
    of the secondary BH. The secondary coordinates are (-19.6, -0.9, 0.1). The lower
    density gas of the gap region has green and blue colors in the
    density maps while the higher density gas of the surrounding
    circum-binary disc is colored in red and yellow. Similarly the
    color scale of the emission line contour levels spans two order of
    magnitudes, uniformly spaced in the logarithm, starting from the
    maximum intensity value in the map with red going to green, blue
    and black. These plots are obtained following the post-process
    analysis procedure described in more detail in the text.}}
\label{fig:contour}
\end{figure*}

In this Section we describe the results from a simulation 
of a close BHB with  $a\simeq 0.01$ pc,  a secondary BH with mass  
$M_2 \simeq 10^8\,\msun$  and mass ratio $q \simeq 0.3$, surrounded
by a coplanar, co-rotating circum-binary disc with a gap.  
In particular, we consider a snapshot where the secondary active 
BH accretes with $f_{\rm Edd}\simeq 0.1$ (we adopt a radiative efficiency of 0.1). 
As discussed in Section \ref{sec:qualitative}, the binary
model implies that  $ R_{\rm{CIV}}\lsim R_{\rm{gap}} <R_{\rm{MgII}} $.
Therefore we expect that the gas in the circum-binary disc 
would give an important contribution to the BELs.

In Figure \ref{fig:contour} we show our results superimposing the
intensity contour plot of the MgII and the CIV lines on the density
map of the selected snapshot. As described in the previous section,
the BELs clouds are distributed on top of the cubic grid adding a
gaseous clump to those illuminated pixels whose hydrogen number
density is $n_{\rm H} \ge 10^8 \,\rm {cm ^{-3}}$. These pixels are
located in the gap region and in a thin layer above the circum-binary
disc midplane, within a radius $r \lsim 2 R_{\rm{MgII}}$
(where $2  R_{\rm{MgII}} \simeq 0.065 \, \rm{pc}$, corresponding to $\simeq 6$ internal unit lengths of the
simulation). As shown in Figure \ref{fig:contour}, the emission line
intensity maps follow the spatial distribution of the gaseous clumps.
We obtain $F_{\rm{MgII}}/F_{\rm{CIV}}\gsim 0.2$ for the emission
coming from the gaseous over-densities. The relative contribution to
the CIV and the MgII lines coming from the gap region is of the order
of $67 \%$ and $34 \%.$

In order to verify that our results are not affected by the details of the
post-process procedure, we perform a second analysis starting from the same
snapshot but considering an isotropic ionizing source. Although the number of
illuminated pixels is $30 \%$ larger in this case, the results are consistent
with the previous analysis. The emission coming from the gap region increases
since the ionizing flux is not zero in the direction perpendicular to the
polar axis. In this case we have $F_{\rm MgII}/F_{\rm CIV} \gsim 0.15$ and the
relative contribution to the CIV and MgII lines from the gaseous clumps in the
central cavity is of the order of $73 \%$ and $25 \%$ respectively.  We point out
that the simplified procedure adopted for the distribution of the
over-densities does not take into account that the formation of the gaseous
clumps might be disfavored in the gap region, characterized by lower
densities and higher temperatures. For this reason $F_{\rm MgII}/F_{\rm CIV}$ could be
higher than what computed with our analysis since we might be overestimating
the contribution to the CIV line coming from the clumps closer to the ionizing
source.

According to Section \ref{sec:qualitative}, we expect to select BHB candidates
with a peculiarly reduced value of $F_{\rm MgII}/F_{\rm CIV}$ when the circum-binary disc
is still too far from the active source and does not give a significant
contribution to the BELs.  As shown in Figure \ref{fig:radii}, $F_{\rm MgII}/F_{\rm CIV}$ 
computed for the BLR around the secondary BH is reduced to a value of $\simeq
0.1$ at an orbital distance of the order of $a \simeq 0.04$ pc, while the inner
edge of the circum-binary disc $R_{\rm gap}$ is located far out at radii $
\simeq 2 R_{\rm{MgII}}$.  In order to quantitatively check such predictions we
perform an additional test using the same snapshot and rescaling the unit
length by a factor of 4 so that the orbital separation of the simulated binary
system is $a \simeq 0.04$ pc. According to eq. 14 of S12, the rescaling keeps
a disc-to-binary mass ratio of a few percent, still consistent with the value
of $1.5 \times 10^{-2}$ assumed in the simulation for $a \simeq 0.01$ pc. Now
we can compare the emission fluxes obtained with the distribution of gaseous
clumps considered for the $a \simeq 0.01$ pc case with the emission fluxes
computed for an equal number of clumps (with the same total mass and
densities) superimposed on the snapshot rescaled at $a \simeq 0.04$ pc. In
particular, the gaseous clumps are added to the rescaled snapshot pixels that are
ionized by the active source and whose number density is $n_{\rm H} > 5 \times
10^6 \,{\rm cm^{-3}}$.
   
With this procedure we find that the contribution to the CIV and the MgII
fluxes from the circum-binary disc, at a binary separation of  $a \simeq 0.04$ pc, is
reduced relative to the case with $a \simeq 0.01$ pc by a factor of $\simeq 10$ and
$\simeq 100$, respectively. This is consistent with our expectations considering
that on average the ionizing flux reaching the high density clumps is
in this case reduced of an order of magnitude. According to the model assumed for the
cloud distribution in the BLR of the secondary BH, at the considered orbital
separation the BELs flux reduction, relative to the fluxes computed at $a_0$,
is of the order of $28 \%$ and $75 \% $ for the CIV and the MgII line,
respectively. According to our results, the contribution from the outer disc
to the CIV line can be neglected while that to the MgII line corresponds to an
increase of $\sim 10 \%$. Even considering this external emission, we obtain 
 $F_{\rm MgII}/F_{\rm CIV} \sim 0.15$ that is still reduced relative to its typical values
for single AGN.  This is consistent with our BHB selection criterion, as discussed in
Paper I, of $F_{\rm MgII}/F_{\rm CIV} \lsim 0.1$ considering both the
uncertainties in the calculations and in the BLR model.

\section{Discussion}\label{sec:discussion}

\begin{figure}
\centering
\includegraphics [scale=0.37] {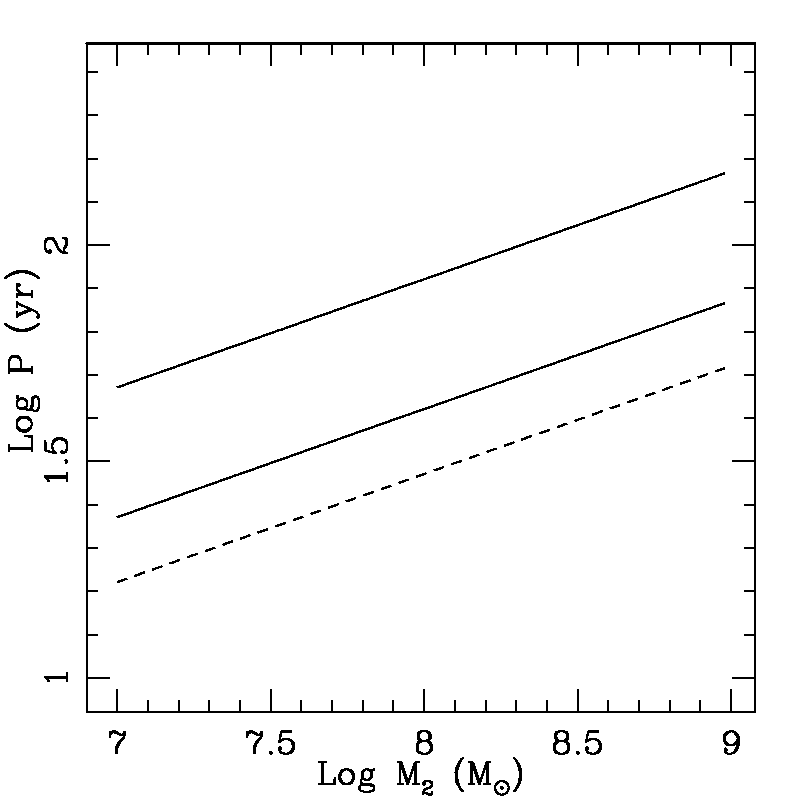}
\includegraphics [scale=0.37] {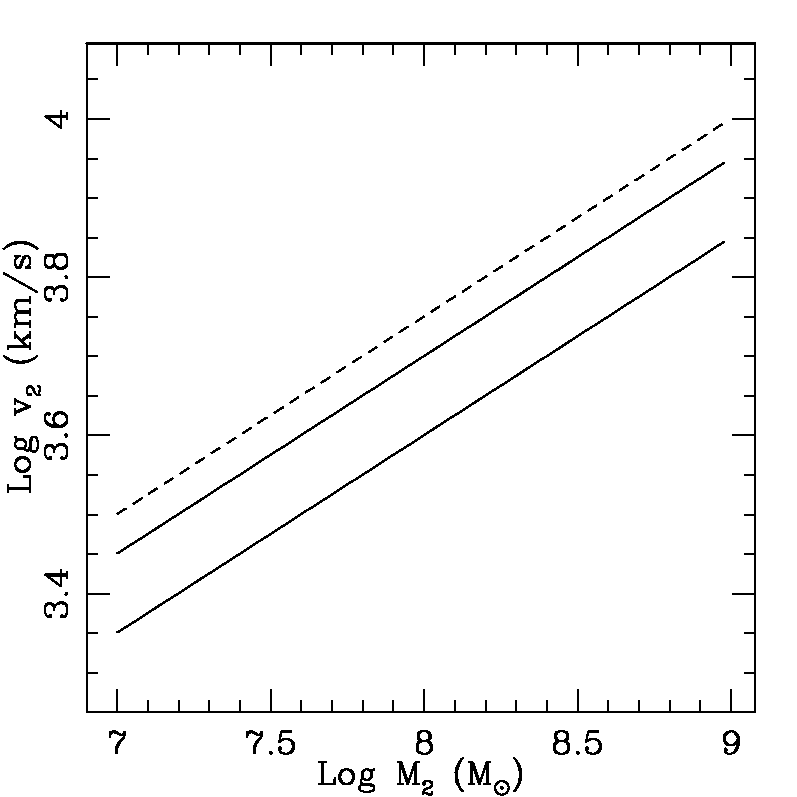}
\caption{\small {Range of orbital parameters of BHB candidates characterized by a
    peculiar value of $F_{\rm MgII}/F_{\rm CIV}$. The
    orbital period $P$ (upper panel) and the orbital velocity of the
    active BH $v_2$ (lower panel) are plotted as function of the mass
    $M_2$ of the active BH. The BH is accreting with
    $f_{\rm Edd}=0.1$, and the binary mass ratio is $q=0.3$. Two
    different models for the BLR bound to the active BH are
    considered. Solid lines refer to the maximum orbital
    period (minimum orbital velocity) for the two considered BLR models.
    Dashed lines indicate the minimum period and the corresponding maximum
    orbital velocity. These parameters are the same for both BLR models
    (see text for details).}}
 
\label{fig:results}
\end{figure}

Figure~\ref{fig:results} shows, as function of the mass of
the secondary active BH, the minimum and maximum values of the orbital
period $P$ (upper panel) and of the orbital velocity $v_2$ (lower
panel) of the active BH relative to the binary center of mass, within
which we expect to select a BHB system from its peculiarly reduced
value of $F_{\rm CIV}/F_{\rm MgII}$.  The corresponding BHB
separations can be inferred given the relation $a \simeq
0.08 \, (M_2/10^8 {\rm{M_{\odot}}})^{1/3} (P/100 \, \rm{yr})^{2/3}$ pc
for a binary mass ratio of 0.3.

Relying on the results discussed in Section~\ref{sec:results},
the limiting
values for $P$ and $v_2$ are obtained requiring that: (i) $R_{\rm gap} \gsim 2
R_{\rm MgII}$, so that the circum-binary disc contributes to the MgII
line at most by $10 \%$ relative to the typical BEL flux of a single
AGN; (ii)  $F_{\rm MgII}/F_{\rm CIV} \lsim 0.15$ even
considering the additional contribution from the gas not bound to the
active BH. We set the upper limit for  $F_{\rm MgII}/F_{\rm CIV}$ to $0.15$ on
the basis of the observed distribution inferred from $\simeq$
6000 quasar spectra in the SDSS archive ($\simeq 90 \%$ of the sources
has a  $F_{\rm MgII}/F_{\rm CIV} > 0.15$, see figure 4 of Paper I).  
Maximum periods (minimum velocities) depend on the way we model
the BLR around the active BH. In the upper panel of Figure~\ref {fig:results} describing 
the run of $P$ with $M_2$, the lower solid line refers to the BLR
model considered in the previous sections, i.e. a uniform density distribution and a
power-law with index -1 for the radial distances distribution (hereon model A). The upper solid line
corresponds to the case of a power-law with index -0.5 for the radial
distribution while the densities are set uniformly in the range
$10^9 {\rm cm^{-3}}\le n_{\rm H} \le 2.5 \times 10^{12}{\rm cm^{-3}}$ (hereon model B). 
In the lower panel for $v_2$ the upper and lower solid lines refer to model A
and model B, respectively. For model B (which has a flatter radial distribution and a lower
upper limit for the density range relative to model A) the  emission coming
from the outer part of the BLR is more relevant and  $F_{\rm MgII}/F_{\rm CIV}$ 
can be significantly reduced also at higher orbital
separations. However in model B we notice that  $F_{\rm MgII}/F_{\rm CIV}$ 
computed at the starting orbital separation, corresponding to
$R_{\rm L}=R_{\rm MgII}$, is lower of a factor of $\sim$ 0.8 because of
the lower density assumed for the BLR clouds. Therefore we consider
the maximum orbital values computed with model B as upper
limits. Moreover for BLR clouds distributed uniformly in both
radius and density, the effect of the flux ratio reduction would be
even stronger than the models considered in the figure.  On the
other hand the BEL fluxes would be so diminished as the orbital
separation shrinks that the contribution from the external gas would
be more effective in increasing  $F_{\rm MgII}/F_{\rm CIV}$  to its standard
values. In this case there would be a narrower range of orbital
parameters suitable to observe a peculiar  $F_{\rm MgII}/F_{\rm CIV}$ because of the
presence of a close BHB.

Solid and dashed lines in Figure \ref{fig:results} scale as $M_2^{1/4}$. 
The scaling can be simply understood considering that the maximum and
minimum orbital separations (thus the maximum and minimum $P$)
correspond to a fixed ratio $R_{\rm L}/R_{\rm BLR}$. This means that $R_{\rm L, max/min} \propto
f_{\rm Edd}^{1/2} M_2^{1/2}$ (see the $R_{\rm{BLR}}$-luminosity
relations reported in section \ref{sec:qualitative}), where the subscript $_{\rm max/min}$
refers to the maximum/minimum value of the parameter. Since $a \propto R_{\rm L}$, we obtain
\begin{equation}
P_{\rm max/min} \propto \left ( \frac{a_{\rm max/min}^{3}}{M_2} \right )^{1/2}\propto \frac{(
  f_{\rm Edd}^{1/2} M_2^{1/2})^{3/2}}{M_2^{1/2}} \propto M_2^{1/4} 
\end{equation}
and 
\begin{equation}
v_{2,{\rm max/min}} \propto \left ( \frac{M_2}{a_{\rm min/max}}
\right )^{1/2} \propto \left ( \frac {M_2}{f_{\rm Edd}^{1/2}
 M_2^{1/2}} \right )^{1/2} \propto M_2^{1/4}
\end{equation}
 As noticed before, we are working under the assumption of a circular binary orbit so that
there is a fixed relation between the Roche radius of the secondary BH
and the binary semi-major axis.  A number of numerical studies showed
that the interaction with a circum-binary disc can bring the
eccentricity of a binary system up to a constant saturation value
which is fairly high, i.e. $e\sim 0.6-0.8$ (e.g. 
Cuadra et al. 2009, Roedig 2011).  The Roche lobe radius is
ill-defined for an eccentric binary.  On an eccentric orbit with
$e\sim 0.6$, the active, secondary BH has an apocenter (pericenter)
$\sim 1.6$ ($\sim 0.4$) times larger (smaller) than a circular
orbit. Under these circumstances, the BLR of the secondary BH may have
no time to re-expand after the truncation at the pericentric passage
due to the characteristics of the periodic inflows of material onto
the BHs. The analysis presented in Section \ref{sec:results} for the
$a\simeq 0.01$ pc case would not be significantly
affected by the precise value of the Roche lobe of the secondary BH. Indeed, 
$R_{\rm{L}} \lsim 0.15 \, R_{\rm{MgII}}$ in both cases
so that the circum-binary disc would still give the major contribution
to the BELs. 

Results are different when most of the emission comes from the material bound to the secondary BH, as in the 
$a\simeq 0.04$ pc case. In this case, if the BLR of the secondary
BH does not expand after the pericentric passage, $F_{\rm MgII}/F_{\rm CIV}$ would
be that computed for a binary with a semimajor axis reduced by a factor $\simeq 2.5$. 
Then, it could be possible to select
binary candidates with longer orbital periods compared to the circular
case. 
If instead the BLR expands to fill $R_{\rm L}$ at the apocentre    
the flux ratio reduction would be less significative. The chance to
select a binary candidate before the circum-binary disc starts to
contribute to the BELs would be lower, and more affected by the
details of the BLR model.

 \section{Summary \& Conclusions}

 In this work we pursue our investigation of new possible observable
 signatures associated with close massive BHBs interacting with a
 circum-binary disc. In Paper I, we proposed a technique based on the
 observation of a peculiarly reduced value of the flux ratio between
 two prominent BELs characterized by different ionization potentials,
 i.e. the CIV and the MgII lines.
 
 In this second work we addressed the robustness of the proposed
 criterion when the presence of the material surrounding the BHB is
 considered. In particular, we aimed at quantifying the effect on
 $F_{\rm{MgII}}/F_{\rm{CIV}}$ of the circum-binary disc and the 
 low density gas in the gap. As in Paper I, we assumed a
 circular BHB with the secondary being the only active BH. We
 presented a qualitative estimate of the relative importance of the
 external contributions to the BEL fluxes at different orbital
 separations. We then tested our results with the post-process
 analysis of a SPH simulation of a BHB embedded in a circum-binary
 disc. The main result of our analysis can be summarized stating that
 we expect to select close massive BHBs because of a peculiarly
 reduced value of $F_{\rm{MgII}}/F_{\rm{CIV}}$ for a limited range of
 sub-pc orbital separations. In this range the main contribution to
 the BELs comes from the tidally perturbed BLR bound to the secondary
 active BH. We notice that, although orbiting at very close
 separations, the BHBs characterized by a low $F_{\rm
   MgII}/F_{\rm CIV}$ are expected to be in the long-lived disc-driven
 phase of the binary orbital migration (e.g. Haiman, Kocsis \& Menou 2009).
 At even closer separations, where the binary lifetimes are shorter
 due to the emission of gravitational waves, the line flux from the
 BLR of the secondary keeps on reducing, while the gas associated
 with the circum-binary disc starts to give a significant contribution
 to the BELs.  This has two main consequences: (i)  $F_{\rm{MgII}}/F_{\rm{CIV}}$ rises up into the typical observed range for
 AGN, and (ii) the wavelengths of the BEL peaks are not expected to
 be related to the orbital motion of the active BH, in agreement with
 Shen \& Loeb (2010) for a different BLR geometry.

\section*{ACKNOWLEDGEMENTS}

\label{lastpage}

\end{document}